\documentclass[conference]{IEEEtran}
\IEEEoverridecommandlockouts

\usepackage{cite}
\usepackage{colortbl}
\usepackage{amsmath,amssymb,amsfonts}
\usepackage{algorithmic}
\usepackage{graphicx}
\usepackage{textcomp}
\usepackage[table]{xcolor}
\def\BibTeX{{\rm B\kern-.05em{\sc i\kern-.025em b}\kern-.08em
    T\kern-.1667em\lower.7ex\hbox{E}\kern-.125emX}}

\usepackage{adjustbox}
\usepackage{balance}
\usepackage{booktabs}
\usepackage{enumitem}
\usepackage{hyperref}
\usepackage{listings}
\usepackage[numbers,sort&compress]{natbib}
\usepackage{siunitx}
\usepackage{subcaption}
\usepackage{tikz}
\usepackage{url}
\usepackage{xspace}

\usepackage[capitalize,sort&compress]{cleveref}

\usetikzlibrary{positioning}

\DeclareSIUnit{\loc}{LOC}

\graphicspath{{./}{img/}}

\definecolor{sbase03}{HTML}{002B36}
\definecolor{sbase02}{HTML}{073642}
\definecolor{sbase01}{HTML}{586E75}
\definecolor{sbase00}{HTML}{657B83}
\definecolor{sbase0}{HTML}{839496}
\definecolor{sbase1}{HTML}{93A1A1}
\definecolor{sbase2}{HTML}{EEE8D5}
\definecolor{sbase3}{HTML}{FDF6E3}
\definecolor{syellow}{HTML}{B58900}
\definecolor{sorange}{HTML}{CB4B16}
\definecolor{sred}{HTML}{DC322F}
\definecolor{smagenta}{HTML}{D33682}
\definecolor{sviolet}{HTML}{6C71C4}
\definecolor{sblue}{HTML}{268BD2}
\definecolor{scyan}{HTML}{2AA198}
\definecolor{sgreen}{HTML}{859900}

\newcommand{\eg}{e.g.,\xspace}
\newcommand{\ie}{i.e.,\xspace}

\newcommand{\papertype}{paper\xspace} %

\newcommand{\tool}{\textsc{SAFE-Deobs}\xspace}
\newcommand{\toolurl}{\url{https://github.com/DSTCyber/safe-deobs}\xspace}
\newcommand{\jsnice}{\textsc{JSNice}\xspace}

\newcommand{\numconstantfoldrules}{\num{59}\xspace}
\newcommand{\toolloc}{\SI{2474}{\loc}\xspace}

\newcommand{\numsamples}{\num{39450}\xspace}
\newcommand{\numvalidsamples}{\num{32341}\xspace}
\newcommand{\numinvalidsamples}{\num{7109}\xspace}
\newcommand{\percentinvalidsamples}{\SI{18.02}{\percent}\xspace}

\newcommand{\numdedupsamples}{\num{28285}\xspace}
\newcommand{\percentdedupsamples}{\SI{71.70}{\percent}\xspace}

\newlist{inlinelist}{enumerate*}{1}
\setlist[inlinelist]{label=(\roman*)}

\lstdefinelanguage{JavaScript}{
  keywords={
    typeof, new, true, false, catch, function, return, null, catch, switch,
    var, if, in, while, do, else, case, break, default, undefined},
  ndkeywords={class, export, boolean, throw, implements, import, this, with},
  sensitive=false,
  comment=[l]{//},
  morecomment=[s]{/*}{*/},
  morestring=[b]',
  morestring=[b]",
}

\def\jsinline{\lstinline[language=JavaScript, basicstyle=\ttfamily, breaklines=false]}

\begin{document}

\title{Optimizing Away JavaScript Obfuscation}

\author{\IEEEauthorblockN{Adrian Herrera}
\IEEEauthorblockA{%
  \textit{Defence Science and Technology Group}
  \\adrian.herrera@dst.defence.gov.au
  }
}

\maketitle

\begin{abstract}
JavaScript is a popular attack vector for releasing malicious payloads on unsuspecting Internet users.
Authors of this malicious JavaScript often employ numerous obfuscation techniques in order to prevent the automatic detection by antivirus and hinder manual analysis by professional malware analysts.
Consequently, this \papertype presents \tool, a JavaScript deobfuscation tool that we have built.
The aim of \tool is to automatically deobfuscate JavaScript malware such that an analyst can more rapidly determine the malicious script's intent.
This is achieved through a number of static analyses, inspired by techniques from compiler theory.
We demonstrate the utility of \tool through a case study on real-world JavaScript malware, and show that it is a useful addition to a malware analyst's toolset.

\end{abstract}


\section{Introduction}\label{sec:introduction}

The past decade has seen JavaScript's popularity steadily increase (and continue to increase): the language is supported by all modern web browsers and used in 95\% of all websites~\cite{W3Techs:2019:JSUsage}; it constantly ranks within the top ten most popular programming languages \cite{StackOverflow:2019:DeveloperSurvey,Tiobe:2019:TiobeIndex,Github:2019:Octoverse}; and thousands of libraries and frameworks have been built on top of it \cite{ModuleCounts:2019:ModuleCounts}.
However, this pervasiveness has a dark side: the ubiquity of JavaScript on the Internet has also made it popular amongst people with malicious intent.
For example, JavaScript is commonly used for gaining initial code execution via a browser or PDF reader vulnerability~\cite{CVE-2019-11707,Jordan:2019:SafePDF}, installing cryptocurrency miners~\cite{TrendMicro:2019:JSMalware}, and in cross-site scripting (XSS) and cross-site request forgery (CSRF) attacks.

The rise of malicious JavaScript (``JavaScript malware'') has resulted in a renewed focus by both antivirus companies and security researchers~\cite{TrendMicro:2019:JSMalware,Xu:2012:PowerJSObfuscation,Lu:2012:AutoDeobJS,Jordan:2019:SafePDF,Fass:2018:JaST,Fass:2019:HideNoSeek}.
In turn, authors of JavaScript malware have increasingly turned to \emph{obfuscation} as a means of
\begin{inlinelist}
  \item hiding from automatic detection by anti-virus products, and
  \item hindering manual analysis by professional malware analysts.
\end{inlinelist}
Thus, tools are required to undo this obfuscation so that JavaScript malware can be detected by antivirus engines and more easily inspected and understood by malware analysts.

We propose and prototype a tool that accomplishes these goals.
Our tool, \tool, is a static analyzer built on top of the SAFE framework~\cite{Lee:2012:Safe1,Park:2017:Safe2}.
It repurposes a number of common compiler optimizations for the purpose of deobfuscating JavaScript malware in order to make it more understandable by subsequent automatic and manual analyses.
This \papertype summarizes our experiences designing, implementing, and evaluating \tool.
Our primary contributions are:

\begin{itemize}
  \item Applying techniques rooted in compiler theory to the task of deobfuscating JavaScript malware;
  \item The design and implementation of \tool, an open-source tool to assist malware analysts to better understand JavaScript malware; and
  \item An evaluation of \tool on a large corpus of real-world JavaScript malware.
\end{itemize}

Unless otherwise stated, all malicious code used in this \papertype is taken from real-world malware.

\section{Background and Related Work}\label{sec:background}

Software obfuscation has many legitimate uses: digital rights management, software diversity (for software protection), and tamper protection, to name a few.
However, software obfuscation is being increasingly co-opted by malware authors to thwart program analysis (both automated and manual).

When discussing JavaScript obfuscation (and obfuscation of other scripting languages; \eg PHP, PowerShell), it is important to differentiate obfuscation from \emph{minification}.
Minification reduces code size by removing unnecessary characters/strings (\eg whitespace and comments) and shortening variable names.
Small code size means less data to download over the Internet, which leads to reduced web page load times.
In contrast, the aim of obfuscation is to make the code difficult to read and understand.
Undoing most minification (commonly referred to as ``beautification'') is trivial---\eg by (re)inserting whitespace---and many tools exist to do this (\eg UglifyJS~\cite{2018:Bazon:UglifyJS}, which we applied to the code in \cref{lst:obfuscated-sample}).
In contrast, deobfuscation often requires advanced program analyses (\eg symbolic execution~\cite{Garba:2019:Saturn} and abstract interpretation~\cite{Preda:2006:OpaquePredicateDetectAI}).
Like \citet{Lu:2012:AutoDeobJS}, we do not consider minification as obfuscation.

\begin{figure}[h]
  \begin{lstlisting}[label=lst:obfuscated-sample,
    caption={A snippet of (beautifed) obfuscated JavaScript.
Comments have been inserted by us.}]
// From 20151226_9474ac02eae3bbe9bcf19d94c8e68a25
var str = "5553515E0A0D0108174A0E05010"/* ... */;
var k6 = ';', e5 = '%
/* 218 more variables */;

h3 += d1;(* \label{line:string-split-start} *)
h3 += x0;
h3 += d7;(* \label{line:string-split-end} *)
/* 215 more operations */
w9(h3);(* \label{line:keyword-sub-call} *)
  \end{lstlisting}
\end{figure}

\Cref{lst:obfuscated-sample} shows a snippet of an obfuscated JavaScript malware sample.
To a human (even a highly-trained malware analyst), it is not immediately obvious what the code in \cref{lst:obfuscated-sample} does: this code defines~222 variables and performs~219 operations on these variables (most of which are not shown, due to space limitations).
An automated tool attempting to either signature the malware or extract features (\eg callback URLs, API calls) will also face difficulties.

Furthermore, JavaScript presents the opportunity to apply obfuscation techniques that are generally unapplicable to compiled languages (\eg C, C++, and Java).
For example, previous work~\cite{Xu:2012:PowerJSObfuscation,Lu:2012:AutoDeobJS} has observed the following obfuscation techniques actively used in the wild:

\begin{description}
  \item[String splitting:] The conversion of a single string into the concatenation of several substrings, as observed on \crefrange{line:string-split-start}{line:string-split-end} in \cref{lst:obfuscated-sample}.
  \item[Keyword substitution:] Storing keywords in variables, such as \jsinline{eval} being stored in \jsinline{w9} on \cref{line:keyword-sub-store} and called via \jsinline{w9} on \cref{line:keyword-sub-call} in \cref{lst:obfuscated-sample}.
  \item[String encoding:] Encode strings so that they are not readable, \eg via escaped ASCII characters, hexadecimal, or Base64 representations.
\end{description}

Finally, JavaScript's ``\textit{quirky semantics}''~\cite{Lee:2012:Safe1} deserve special mention.
These include:

\begin{description}
  \item[Variable hoisting:] JavaScript allows variable declarations \emph{after} the variable has been used.
These variable declarations are moved (``hoisted'') to the top of the scope in which they are used.
  \item[\texttt{with} scoping:] Extends the scope of the current statement, which is specially noted in the Mozilla developer documentation ``\textit{as it may be the source of confusing bugs}''~\cite{Mozilla:2020:JSWith}.
  \item[\texttt{eval}:] Dynamically executes JavaScript code and is a known security risk~\cite{Mozilla:2020:JSEval}.
\end{description}

These semantics can be abused (by both malicious and benign code alike) to make both static analysis and deobfuscation difficult~\cite{Lee:2012:Safe1,Jordan:2019:SafePDF}.

To this end, existing work on JavaScript (de)obfuscation has mostly focused on detecting (and preventing) malicious JavaScript, often using machine learning~\cite{Rieck:2010:Cujo,Curtsinger:2011:Zozzle,Fass:2018:JaST,Fass:2019:JStap,Stokes:2019:ScriptNet} or program analysis~\cite{Jordan:2019:SafePDF} techniques.
Machine learning approaches (\eg random forests, support vector machines) have been shown to be effective when combined with \emph{semantic} features (\eg control flow and program dependency graphs, as used by \textsc{JStap}~\cite{Fass:2019:JStap}).
Similarly, program analysis techniques (such as abstract interpretation) that operate on (an abstraction of) JavaScript's semantics have also been effective at detecting JavaScript malware~\cite{Jordan:2019:SafePDF}.
This is likely due to the fact that obfuscation must maintain the original code's \emph{meaning} (\ie its semantics), while syntactic features are easier to manipulate.
However, this focus on \emph{detection} has traditionally eschewed \emph{readability}: the focus of deobfuscation.
Our work therefore complements much of this existing research.

\subsection{Related Work}\label{sec:related-work}

\Citet{Lu:2012:AutoDeobJS} combine dynamic analysis (to capture a JavaScript execution trace) and static analysis (backward slicing of the execution trace to create a simplified Abstract Syntax Tree) to produce \emph{observationally equivalent} output JavaScript.
Dynamic analysis allows for code ``hidden'' in an \jsinline{eval} call to be analyzed and deobfuscated.
However, dynamic analysis can also be thwarted by environmental checks that may not be satisfied when the malware is under analysis.
Their prototype is also not publicly available.

\jsnice uses ``big code'' and machine learning to predict program properties (\eg variable names and type annotations) for JavaScript code~\cite{Raychev:2015:JSNice}.
While not open-source, the \jsnice authors have created a website\footnote{\url{http://www.jsnice.org/}} that promises to ``\textit{make even obfuscated JavaScript code [uploaded to the website] readable}''.
Unfortunately, uploading the code from \cref{lst:obfuscated-sample} resulted in the code in \cref{lst:jsnice-deobfuscated-sample}: while the type annotations are accurate, this code is no-more readable than that in \cref{lst:obfuscated-sample}.

\begin{figure}[h]
  \begin{lstlisting}[label=lst:jsnice-deobfuscated-sample,
    caption={A snippet of the JavaScript from \cref{lst:obfuscated-sample} ``deobfuscated'' by \jsnice.
Comments enclosed within \jsinline{/** */} (\ie the type annotations) were generated by \jsnice.
Other comments have been inserted by us.}]
'use strict';
/** @type {string} */
var str = "5553515E0A0D0108174A0E05010"/* ... */;
/** @type {string} */
var k6 = ";";
/** @type {string} */
var e5 = '%
/** @type {string} */
var h3 = "";
/** @type {function(string): *} */
var w9 = eval;
/* 218 more variables */

/** @type {string} */
h3 = h3 + d1;
/** @type {string} */
h3 = h3 + x0;
/** @type {string} */
h3 = h3 + d7;
/* 215 more operations */
w9(h3);
  \end{lstlisting}
\end{figure}

Other tools, such as ``deobfuscate javascript''\footnote{\url{http://deobfuscatejavascript.com/}} perform deobfuscation by intercepting calls to \jsinline{eval} and \jsinline{write}.
This approach is able to successfully deobfuscate the sample in \cref{lst:obfuscated-sample} (due to the \jsinline{eval} substitution on \cref{line:keyword-sub-call}).
However, as noted on their website: ``\textit{some malicious scripts may not employ these functions and may therefore infect your browser}''.

This opens the door for a new, open-source tool for JavaScript malware deobfuscation.
To this end, the following sections discuss the design and implementation of our contribution to this field.

\section{Design and Implementation}\label{sec:design-impl}

\begin{figure}
  \centering
  \includegraphics[width=0.8\columnwidth]{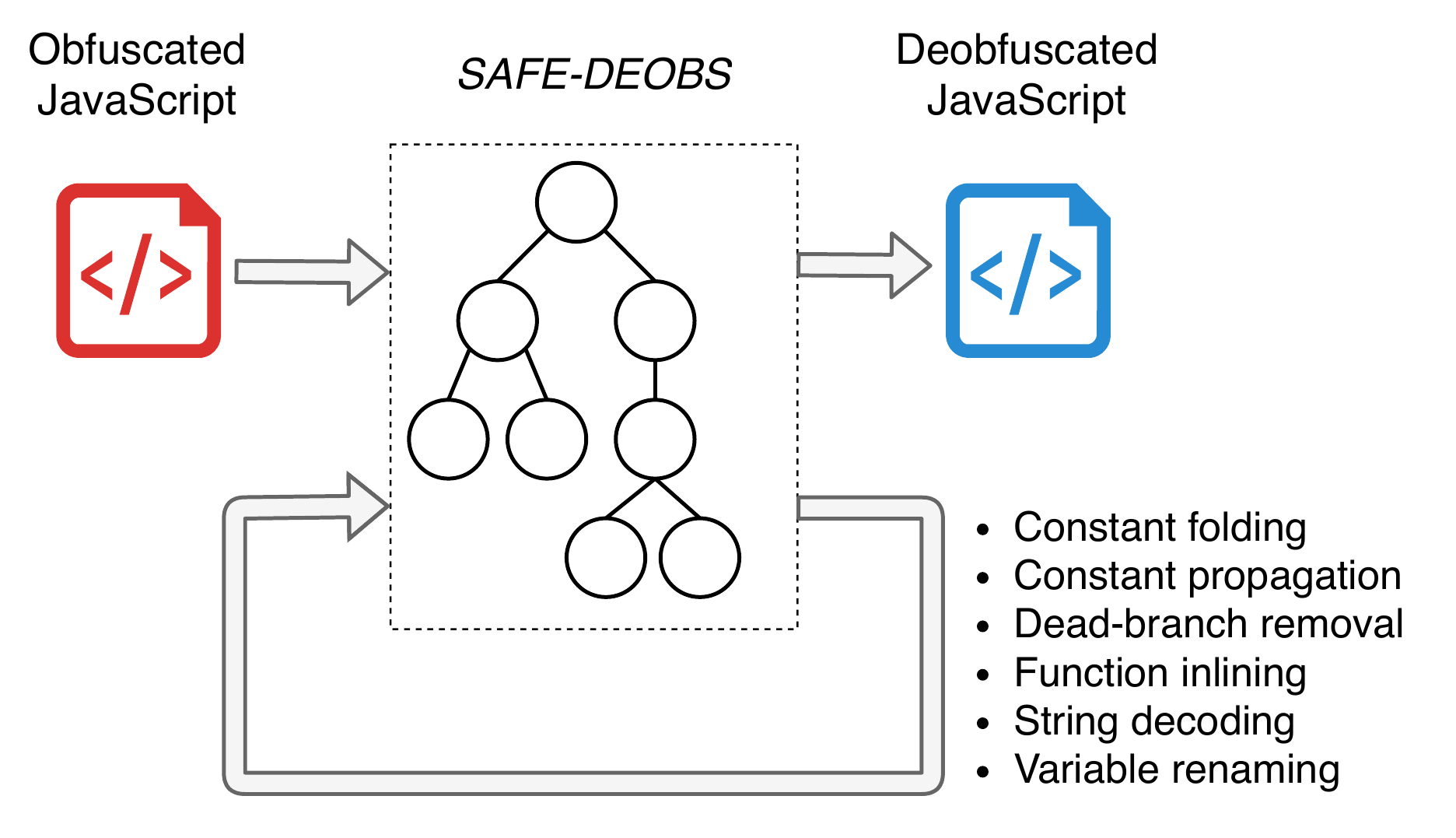}
  \caption{The \tool workflow.}
  \label{fig:workflow}
\end{figure}

The \tool  workflow is illustrated in \cref{fig:workflow}.
The obfuscated JavaScript is parsed into an Abstract Syntax Tree (AST), upon which a number of deobfuscation passes are performed (much like a compiler performing a set of optimization passes).
These passes are continuously applied until a fixed point is reached, at which point the AST is serialized back into JavaScript source code.
The set of deobfuscation passes that we developed (listed in \cref{fig:workflow}) will be described in \cref{sec:deobfuscation-passes}.
But first, the JavaScript must be transformed into a form amendable to analysis.

\subsection{Preprocessing}\label{sec:preprocessing}

We reuse existing tools to parse and preprocess the input JavaScript.
Specifically, we use the Scalable Analysis Framework for ECMAScript (SAFE~v2.0)~\cite{Lee:2012:Safe1,Park:2017:Safe2} to parse JavaScript into an AST that can be further analyzed.

SAFE is a scalable analysis for JavaScript\footnote{We use ECMAScript and JavaScript interchangeably throughout this \papertype.} that provides different levels of intermediate representations (IR).
We selected SAFE because it is open-source and ``\textit{especially designed as a playground for advanced research in JavaScript web applications}''~\cite{Park:2017:Safe2}.
Furthermore, SAFE is primarily written in Scala, a functional programming language.
Features standard in functional programming languages---such as pattern-matching and support for efficient recursion---lend themselves well to writing analyses/transformations that operate on trees (\ie ASTs).

While SAFE provides three levels of IR, we only use the lowest level: the AST.
The higher levels (the Intermediate Representation and Control Flow Graph) are not used because they only support one-way translation: from lower levels up to higher levels.
As the higher IRs are aimed at analysis, rather than transformation, there is no mechanism to translate higher IRs back to JavaScript.

Finally, we make use of two features in SAFE's AST translator that simplify further analysis: the \texttt{Hoister} and \texttt{WithRewriter}.
The \texttt{Hoister} lifts variable declarations so that they appear at the top of the current scope, separating declarations from initializations.
The \texttt{WithRewriter} conservatively eliminates \jsinline{with} statements such that lexical scoping remains valid.
After this, the AST is ready for deobfuscation.

\subsection{Deobfuscation Passes}\label{sec:deobfuscation-passes}

After preprocessing, \tool performs a number of ``deobfuscation passes'' (``phases'' in SAFE parlance) on the AST until a fixed point is reached.
These phases include:
\begin{inlinelist}
  \item constant folding;
  \item constant propagation;
  \item dead-branch removal;
  \item function inlining;
  \item string decoding; and
  \item variable renaming,
\end{inlinelist}
totalling~\toolloc of Scala.
The first four phases should be familiar to those with an understanding of common compiler optimizations, while the last two are specific to scripting languages.
We describe each of these phases in the following sections.

\subsubsection{Constant Folding}\label{sec:constant-folding}

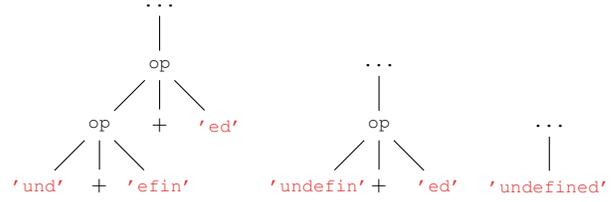
\begin{figure}
  \centering
  \begin{subfigure}[t]{\columnwidth}
    \centering
    \begin{lstlisting}[numbers=none]
if (typeof ifopracxa == 'und' + 'efin' + 'ed')
    \end{lstlisting}
    \caption{The original JavaScript code.}
    \label{lst:unfolded-code}
    \vspace{1em}
  \end{subfigure}
  \begin{subfigure}[t]{\columnwidth}
    \centering
    \begin{tikzpicture}[font=\scriptsize\ttfamily, level distance=8mm, sibling distance=8mm]
       \node  {\ldots}
         child { node {op}
           child { node {op}
             child { node (str start) {\color{sred}{'und'}} }
             child { node {$+$} }
             child { node (str end) {\color{sred}{'efin'}} }
           }
           child { node {$+$} }
           child { node {\color{sred}{'ed'}} }
         };
    \end{tikzpicture}
    \begin{tikzpicture}[font=\scriptsize\ttfamily, level distance=8mm, sibling distance=8mm]
       \node  {\ldots}
         child { node {op}
           child { node (str start) {\color{sred}{'undefin'}} }
           child { node {$+$} }
           child { node (str end) {\color{sred}{'ed'}} }
         };
    \end{tikzpicture}
    \begin{tikzpicture}[font=\scriptsize\ttfamily, level distance=8mm, sibling distance=8mm]
      \node  {\ldots}
        child { node {\color{sred}{'undefined'}} };
    \end{tikzpicture}
    \caption{The AST as constant folding is applied (from left to right).}
    \label{fig:constant-fold-ast}
  \end{subfigure}
  \caption{Constant folding example.}
  \label{fig:constant-folding}
\end{figure}

Constant folding aims to recognize and evaluate constant expressions.
This process is illustrated in \cref{fig:constant-folding}, where \emph{string splitting} has been applied to a string literal (\cref{lst:unfolded-code}) in an \jsinline{if} condition.
The original string can be recovered by traversing the AST (\cref{fig:constant-fold-ast}) and rewriting nodes where an arithmetic operation occurs on two constant nodes.

SAFE's AST representation and Scala's pattern-matching feature make this straightforward to implement: \cref{lst:scala-constant-fold} gives an example of one such rule (the concatenation of strings with integers).
While we have implemented~\numconstantfoldrules rules, JavaScript's quirky semantics (\cref{sec:background}) means that corner-cases may remain unhandled.
Fortunately, these rules are straightforward to extend.

\begin{figure}[h]
  \centering
  \begin{lstlisting}[language=scala, label=lst:scala-constant-fold,
      caption={An example of a constant folding rewrite rule.
This rule matches on a string concatenated with an integer, resulting in a single string literal (according to JavaScript's semantics).}]
case InfixOpApp(StringLiteral(quote, str, false),
                Op("+"), IntLiteral(int)) =>
  StringLiteral(quote, s"${str}${int}", false)
  \end{lstlisting}
\end{figure}

\subsubsection{Constant Propagation}\label{sec:constant-propagation}

Constant propagation is the substitution of known literal values into expressions.
Unlike constant folding (\cref{sec:constant-folding}), constant propagation requires
\begin{inlinelist}
  \item maintaining state while the AST is traversed, and
  \item multiple traversals of the AST.
\end{inlinelist}

\begin{figure}
  \centering
  \begin{tikzpicture}[outer sep=0pt, node distance=15pt and 8pt, every node/.style={scale=0.75}]
    \node (top) {$\top$};

    \node[below=of top] (zero) {\jsinline{0}};
    \node[right=of zero] (str) {\jsinline{"aabb"}};
    \node[right=of str] (false) {\jsinline{false}};
    \node[right=of false] (inf) {\ldots};

    \node[left=of zero] (neg one point one) {\jsinline{-1.1}};
    \node[left=of neg one point one] (undefined) {\jsinline{undefined}};
    \node[left=of undefined] (neg inf) {\ldots};

    \node[below=of zero] (bot) {$\bot$};

    \draw (bot) -- (neg inf) -- (top);
    \draw (bot) -- (undefined) -- (top);
    \draw (bot) -- (neg one point one) -- (top);
    \draw (bot) -- (zero) -- (top);
    \draw (bot) -- (str) -- (top);
    \draw (bot) -- (false) -- (top);
    \draw (bot) -- (inf) -- (top);
  \end{tikzpicture}
  \caption{The constant propagation lattice.}
  \label{fig:constant-propagation-lattice}
\end{figure}
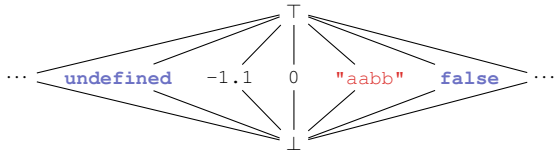

The state of a variable's constness must be tracked during AST traversal: value substitution can no longer occur once a variable's constness can no longer be guaranteed.
We implement this as an abstract interpretation over the three-level lattice typically used for constant propagation~\cite{Wegman:1991:ConstPropCondBranches} (\cref{fig:constant-propagation-lattice}).
This lattice contains an infinite number of middle elements, representing constant values.
A variable goes to~$\top$ once its value is no longer constant.

Once constants have been propagated through the AST, the AST is traversed again to remove redundant variable assignments.

\subsubsection{Dead-branch Removal}\label{sec:dead-branch-removal}
Dead code is often inserted to confuse and distract a malware analyst.
This dead code is often revealed by other deobfuscation phases (in particular, constant folding and propagation).
The simplest form of dead code removal is when a constant (\ie \texttt{Literal} AST node) appears in an \jsinline{if} condition.
In this instance, we can statically determine which branch will \emph{always} execute, and remove the other branch.
We apply the same technique to \jsinline{switch} statements, removing unreachable \jsinline{case}s.

\subsubsection{Function Inlining}\label{sec:function-inlining}

Function inlining expands trivial function calls, where the complexity of the analysis determines what is ``trivial''.
Like the dead code discussed in \cref{sec:dead-branch-removal}, obfuscation may introduce
\begin{inlinelist}
  \item functions that are never called, and/or
  \item functions that perform trivial operations, but nevertheless adds to an analyst's cognitive load.
\end{inlinelist}

Our prototype inlines functions where the function's AST consists of a single \texttt{Return} statement that returns a \texttt{Literal} expression.
Examples of such functions are given in \cref{lst:function-inline-examples}.
The first function is trivial to inline.
The second function (\jsinline{Ph}) becomes inlinable after string decoding and constant propagation (\cref{sec:string-decoding,sec:constant-propagation} respectively).
Again, Scala's pattern-matching feature allows for a concise rule to find such functions (\cref{lst:scala-function-inline}).

\begin{figure}[h]
  \begin{lstlisting}[label=lst:function-inline-examples,
    caption={Examples of inlinable functions.}]
// From 20170124_a0b2eeedbc9c6187927e32645700d1d2
function zdykuvpobrenusdegvusipasad/*- ... -*/() {
  return [ /*- ... -*/ , "ing", /*- ... -*/];
}

// From 20190808_536f24111b28ff9febcdaef4ceb47adb
function Ph() {
  var fHC=String.fromCharCode(6688/88+0);(* \label{line:fromcharcode-start} *)
  nDO = fHC + String.fromCharCode(2600/52-0);
  /* 7 more operations */
  oOw = Oy + String.fromCharCode(16*5);(* \label{line:fromcharcode-end} *)
  return oOw;
}
  \end{lstlisting}
\end{figure}

\begin{figure}[h]
  \begin{lstlisting}[language=scala, label=lst:scala-function-inline,
    caption={Pattern-matching rules for an inlinable function.}]
// Match the last statement in the function body
// Precondition: funcBody contains a single stmt
funcBody.last match {
  // Function returns a literal expression
  case Return(Some(lit: Literal)) => Some(lit)
  // Function returns nothing
  case Return(None) => EmptyExpr
  // Function returns one of its parameters
  case Return(Some(VarRef(id))
    if params.exist(_ == id) =>
      // Return the param with the given id
  // Cannot inline
  case _ => None
}
  \end{lstlisting}
\end{figure}

Once inlinable functions are found (by traversing the AST), the AST is traversed (again) so that all \texttt{FunApp} (\ie function application) nodes that call an inlinable function are replaced with the \texttt{Literal} expression returned by the function.
For example, calls to the first function in \cref{lst:function-inline-examples} are replaced with the returned array literal, while calls to the second function are replaced with the literal value in \jsinline{oOw} (after string decoding and constant propagation have been applied).

\subsubsection{String Decoding}\label{sec:string-decoding}

As discussed in \cref{sec:background}, strings can be encoded such that they are unreadable by most analysts.
Common string encoding schemes include hexadecimal (typically found in string literals, \eg \jsinline{'\x68\x65\x6c\x6c\x6f'}) and unicode (commonly decoded using the \jsinline{String.fromCharCode} function, as used in \crefrange{line:fromcharcode-start}{line:fromcharcode-end} in \cref{lst:function-inline-examples}).
These encoding patterns are straightforward to find---by looking for \texttt{StringLiteral} AST nodes that contain hexadecimal-encoded strings and calls to \jsinline{String.fromCharCode}, respectively---and rewrite.

The encoding schemes thus far have all utilized language features built into the JavaScript language.
Of course, malware authors are free to implement other encoding (\eg Base64) or encryption (\eg RC4) schemes.
While statically detecting these encoding/encryption schemes is impossible in the general-case, it may still be possible to employ heuristics to detect such functionality.
While our prototype does not support this, SAFE provides an ideal environment to develop and experiment with such heuristics.

\subsubsection{Variable Renaming}\label{sec:variable-renaming}

Finally, variables (and functions) can be given complex, confusing and/or similar names to increase an analyst's cognitive load.
This is demonstrated in \cref{lst:obfuscated-variable-names}.

\begin{figure}
  \centering
  \begin{subfigure}[t]{0.47\columnwidth}
    \centering
    \begin{lstlisting}
var 1I11II;
if (III1II() == 1I11I1)
  11I11I();

1I11II = 11I1I(1I11I1);
    \end{lstlisting}
    \caption{The original code.}
    \label{lst:obfuscated-variable-names}
  \end{subfigure}
  \hfill
  \begin{subfigure}[t]{0.46\columnwidth}
    \centering
    \begin{lstlisting}
var dog; // 1I11II
if (cat() == parrot)
  lion();

dog = tiger(parrot);
    \end{lstlisting}
    \caption{After variable renaming.}
    \label{lst:renamed-variables}
  \end{subfigure}
  \caption{An example of variable/function renaming.}
  \label{fig:variable-renaming}
\end{figure}

Variable names that share common prefixes complicate variable renaming via regular expressions (depending on the order in which variables are renamed).
Therefore, we developed an optional\footnote{Occasionally, malware will declare variables such as \jsinline{var exploitation}, which are useful names.} SAFE phase that renames all variables to animal names.
Animals are deterministically selected so that repeated deobfuscation of the same sample produces the same result.
The original variable names are placed alongside renamed variable definitions, in case analysts are required to refer back to the original malware sample, as in \cref{lst:renamed-variables}.

\section{Evaluation}\label{sec:evaluation}

Here we present
\begin{inlinelist}
  \item a case study on a particular malware sample (\cref{sec:case-study}); and
  \item a high-level analysis over~\numsamples malware samples (\cref{sec:generalizability}).
\end{inlinelist}
Both of these discussion will use the open-source dataset from Hynek Petrak~\cite{HynekPetrak:2019:JSMalwareCollection}.

\subsection{Case Study}\label{sec:case-study}

This case study is based on the JavaScript malware sample \texttt{20170110\_9330ee612a9027120543d6cd601cda83}, which is publicly available from our dataset.%

This particular sample has not been minified and consists of~\SI{475}{\loc}, contains~14 functions, and defines~214 variables.
This sample makes for an interesting case study because it is one of the few samples that does not use \jsinline{eval}, and ``deobfuscate javascript'' (\cref{sec:background}) is therefore unable to deobfuscate it (due to its reliance on hooking \jsinline{eval}).
Interestingly, \jsnice is unable to infer any of the~14 functions' return types, despite all of these functions being inlinable (according to our inline rules, described in \cref{sec:function-inlining}) and therefore relatively straightforward to analyze.

\begin{figure}
  \centering
  \begin{lstlisting}[label=lst:deob-case-study-1,
         caption={Case study sample after deobfuscation.}]
var lion; // edeb
var hamster; // uvacdykadq
var chinchilla; // cqorobcit

lion = WScript;
hamster = typeof window == "undefined";(* \label{line:dom-window} *)
{
  chinchilla = lion.CreateObject('WScript.Shell');
  if (hamster) {
    chinchilla["run"]('cmd.exe /c \"powershell  $ojogo=\'^dimas.top\';$hetfo=\'^-Scope  Pr\';$pobbi=\'^,$path); \';$innypu=\'^ocess; $p\';$monsucm=\'^y Bypass \';$binkucb=\'^h\';$ykpyffy=\'^Start-Pro\';$ykjygr=\'^:temp+\'\'\b\';$uzmez=\'^e\'\');(New-\';$xzymo=\'^Set-Execu\';$ulirgo=\'^tp://laro\';$eqtem=\'^ath=($env\';$evyvz=\'^).Downloa\';$ogxow=\'^Webclient\';$utkyjv=\'^/777.exe\'\'\';$gsydibv=\'^tionPolic\';$upoh=\'^stem.Net.\';$zceqmi=\'^Object Sy\';$cepsuhm=\'^ipbafa.ex\';$qfyzko=\'^dFile(\'\'ht\';$awysqe=\'^cess $pat\'; Invoke-Expression ($xzymo+$gsydibv+$monsucm+$hetfo+$innypu+$eqtem+$ykjygr+$cepsuhm+$uzmez+$zceqmi+$upoh+$ogxow+$evyvz+$qfyzko+$ulirgo+$ojogo+$utkyjv+$pobbi+$ykpyffy+$awysqe+$binkucb);\"', 0);(* \label{line:obfuscated-powershell} *)
  }
}
  \end{lstlisting}
\end{figure}

\Cref{lst:deob-case-study-1} shows the deobfuscated sample.
All~14 functions have been inlined and~211 variables have been eliminated through a repeated combination of constant folding and propagation (a fixed point was reached after four iterations).
The number of lines has been reduced by~\SI{97}{\percent} to~\SI{12}{\loc}.

Unfortunately, \jsinline{hamster} (\cref{line:dom-window}) remains because we do not model the Document Object Model (DOM, of which the \jsinline{window} object is an element of).
Nevertheless, it is now much easier to reason about the sample's behavior.
Alas, this behavior primarily consists of executing a string of obfuscated PowerShell (\cref{line:obfuscated-powershell}).
Clearly, \tool would benefit from integration with other malware analysis/deobfuscation tools.

\subsection{Generalizability}\label{sec:generalizability}

We examined all~\numsamples malware samples in our dataset to obtain a high-level understanding of our tool's efficacy.
First, we removed~\numinvalidsamples invalid samples (\percentinvalidsamples of the dataset) that escomplex~\cite{2017:Booth:escomplex}---a tool for performing software complexity analysis on JavaScript ASTs---failed to parse.
Second, we ``normalized'' the dataset by running all remaining~\numvalidsamples through SAFE's \texttt{astRewrite} phase, which hoists variable definitions, rewrites \jsinline{with} statements, and beautifies the code (\ie undoes minification).
This allowed us to remove duplicate samples (identified by SHA512 checksum), leaving~\numdedupsamples samples (\percentdedupsamples of the original dataset).

Finally, we ran \tool over the~\numdedupsamples deduplicated samples and used escomplex to compare common software complexity metrics before and after deobfuscation.
These complexity metrics include:
\begin{inlinelist}
  \item physical lines of code (LOC);
  \item number of functions;
  \item cyclomatic complexity~\cite{McCabe:1976:Cyclomatic}; and
  \item Halstead length~\cite{Halstead:1977:Halstead}.
\end{inlinelist}
The results are presented in \cref{tab:deobfuscation-stats}.

\begin{table}
  \centering
  \caption{Complexity metrics before and after deobfuscation (using the normalized, deduplicated samples from our dataset).}
  \label{tab:deobfuscation-stats}

  \begin{adjustbox}{width=0.99\linewidth}
  \rowcolors{2}{white}{gray!25}
  \begin{tabular}{lrrr}
    \toprule
    Metric                     & Before         & After          & \% decrease \\
    \midrule
    Total physical LOC         & \num{46724630} & \num{45491768} & \num{2.64} \\
    Total num.\ functions      & \num{324441}   & \num{241091}   & \num{25.69} \\
    Mean cyclomatic complexity & \num{10.58}    & \num{8.68}     & \num{17.96} \\
    Mean Halstead length       & \num{5994.62}  & \num{4297.03}  & \num{28.31} \\
    \bottomrule
  \end{tabular}
  \end{adjustbox}
\end{table}

We used both escomplex's report and manual inspection to verify deobfuscation correctness.
Unfortunately, JavaScript malware is difficult to verify via dynamic behavioral analysis because:
\begin{inlinelist}
  \item the malware's output may vary depending on its intent (\eg downloading a second-stage implant, exploiting a vulnerability) and may not be readily apparent;
  \item the malware may require complex ``trigger conditions''~\cite{Brumley:2008:TriggerBasedMalware} to activate the intended behavior; and
  \item the malware may target a particular browser.
\end{inlinelist}
Nevertheless, we found that \tool successfully processed our malware corpus and greatly reduced the complexity of the code contained within.

\section{Conclusion}\label{sec:conclusion}

In this \papertype we present \tool, a static analyzer for deobfuscating JavaScript malware.
While none of the techniques that we propose are particularly novel in their own right---indeed, \citet{Garba:2019:Saturn} also proposed ``\textit{deobfuscation by optimization}''---little work has been published on applying these techniques to JavaScript malware.
We have demonstrated \tool' utility by applying it to a large corpus of real-world malware, and shown that it makes for a useful addition to a malware analyst's toolset.
\tool is open-source and available to malware analysts at \toolurl.

\balance
\bibliographystyle{IEEEtranN}
\bibliography{safe-deobs}

\end{document}